%% file: janka_main.tex
\documentclass[multphys,vecphys]{svmult}

\usepackage{makeidx}     
\usepackage{graphicx}    
\usepackage{multicol}    
\usepackage{natbib}
\usepackage{amssymb}
\usepackage{amsmath}
\usepackage{times}
\usepackage{epsfig}
\usepackage[figuresright]{rotating}

\makeindex             
\begin{document}

\title*{Core-Collapse Supernovae at the Threshold}
\titlerunning{Core-Collapse Supernovae}
\author{H.-Th.~Janka, R.~Buras, K.~Kifonidis, A.~Marek, and M.~Rampp}
\institute{Max-Planck-Institut f\"ur Astrophysik, 
Postfach 1317, D-85741 Garching, Germany\\
\texttt{thj@mpa-garching.mpg.de}}
%
%
\maketitle

\begin{abstract}
Recent progress in modeling core-collapse supernovae
is summarized and set in perspective. Two-dimensional
simulations with state-of-the-art treatment of neutrino
transport still fail to produce powerful explosions, but 
evidence is presented that they are very close to
success.
\end{abstract}

\section{Aiming high}
\label{janka_sec:1}
Despite of still bothering uncertainties and ongoing
controversy, the convectively supported neutrino-heating 
mechanism~\cite{thj_herben92} must be considered as a 
promising way to explain supernova explosions of massive stars.
Neutrinos drive the
evolution of the collapsing stellar core and of the forming   
neutron star and dominate the event energetically
by carrying
away about 99\% of the gravitational binding energy of the 
compact remnant. A detailed description of their processes 
in models which couple (relativistic)
hydrodynamics and accurate neutrino transport is therefore
indispensable for making progress towards an understanding
of the remnant-progenitor connection, supernova energetics, 
explosion asymmetries, pulsar kicks, nucleosynthesis, and
observable neutrino and gravitational-wave signals. It is
a necessary ingredient in any calculation which claims a
higher degree of realism.

\section{Stepping forward}
\label{janka_sec:2}
Successful simulations of neutrino-driven supernova explosions
have so far either employed special,
usually controversial, assumptions about the physics at
neutron star conditions or have made use of crude 
approximations in the neutrino transport. They are contrasted
by simulations with widely accepted microphysics and
an increasing sophistication of the transport treatment,
which have not been able to produce explosions.

It must be stressed, however, that these
simulations do not conflict with each other.
They were performed with largely different numerical
descriptions and the discrepant results 
simply demonstrate the sensitivity of the
delayed explosion mechanism to variations at the level of
the different approaches.

\subsection{Successful explosions on the one hand...}
Wilson and collaborators~\cite{thj_wiletal} found explosions 
in one-dimensional simulations by assuming
that neutron finger convection below the neutrinosphere boosts
the neutrino emission from the nascent neutron star and thus
increases the neutrino heating behind the stalled supernova
shock. Neutron finger convection, however, requires a faster
exchange of energy than lepton number between fluid
elements, an assumption that could not be confirmed by 
detailed analysis of the multi-flavor neutrino 
transport~\cite{thj_brudin96}.
Another ingredient to the energetic explosions of Wilson's group
is a nuclear equation of state (EoS) which yields high 
neutron star temperatures and $\nu_e$ luminosities because
of the formation of a pion condensate at rather low 
densities~\cite{thj_maytav93}. The adopted
dispersion relation of the pions in dense matter, however, is
not supported by accepted nuclear physics.

In the early 1990's it was recognized that violent 
convective overturn between the neutrinosphere
and the supernova shock is
helpful for the neutrino-heating mechanism and a possible
origin of the anisotropies and large-scale mixing observed in 
SN~1987A~\cite{thj_herben92}. 
Two-dimensional hydrodynamic models~\cite{thj_snconv}
and, more recently, 3D simulations~\cite{thj_frywar02} 
that take this
effect into account produced explosions, but used grey 
(i.e., spectrally averaged), flux-limited diffusion for
decribing the neutrino transport, an approximation which 
fell much behind the elaborate multi-group diffusion that
had been applied by Bruenn in spherical symmetry~\cite{thj_bru85}.

\subsection{...and failures on the other}
Bruenn, using standard microphysics and a sophisticated
multi-group flux-limited diffusion treatment of neutrino
transport, could never confirm explosions in
one-dimensional simulations~\cite{thj_bru93}. But there was hope 
that an even better description of the transport
might bring success.

A new level of accuracy has indeed been reached 
with the use of solvers for the Boltzmann transport equation.
Employing different numerical techniques, they were only
recently applied in time-dependent hydrodynamic 
simulations of spherical stellar core collapse with 
Newtonian gravity~\cite{thj_bolnewt}, approximate treatment of 
relativistic effects~\cite{thj_ramjan02}, 
and general relativity~\cite{thj_lieetal01}.
These simulations, all performed with the EoS of Lattimer 
\& Swesty~\cite{thj_latswe91}, agree that neither prompt
explosions by the hydrodynamic bounce-shock mechanism,
nor delayed, neutrino-driven explosions could be obtained 
without the help of
convection, not even with the best available treatment of
the neutrino physics and general relativity. 

Mezzacappa et al.~\cite{thj_mezetal98} 
also expressed concerns that the success of multi-dimen\-sional
calculations~\cite{thj_snconv,thj_frywar02} might 
disappear once the neutrino transport is improved to the
sophistication reached in 1D models. 
They demonstrated this by mapping
transport results from 1D supernova models to 2D hydrodynamics.
The lacking self-consistency of this approach, however, was
an obvious weakness of the argument.

\section{Pushing the limits}
\label{janka_sec:3}
In this situation the core-collapse group at Garching 
has advanced to the next level of improvements in supernova 
modeling. To this end we have generalized our 1D 
neutrino-hydrodynamics code (VERTEX; \cite{thj_ramjan02}) 
for performing
multi-dimensional supernova simulations with a state-of-the-art
treatment of neutrino transport and neutrino-matter
interactions, calling the extended code version 
MuDBaTH~\cite{thj_mudbath}.

\subsection{A new tool...}

The hydrodynamics part of the program is based on the
PROMETHEUS code, which is an Eulerian finite-volume method for
second-order, time-explicit integration of the hydrodynamics
equations. It employs a Riemann solver for 
high-resolution shock capturing, a consistent multi-fluid
advection scheme, and general relativistic corrections to the 
gravitational potential. ``Odd-even decoupling'' at
strong shocks is avoided by an HLLE solver. More details about 
technical aspects and corresponding references can be found
in Refs.~\cite{thj_ramjan02,thj_mudbath}.

The hydro routine is linked to a code which solves the 
multi-frequency transport problem for neutrinos and
antineutrinos of all flavors by closing the set of
moment equations for particle number, energy and momentum
with a variable Eddington factor that is 
computed from a model Boltzmann equation. The transport is 
done in a time-implicit way and takes into account
moving medium effects and general relativistic redshift
and time dilation. Transport and hydro components are
joined by operator-splitting. The multi-dimensional
version of the code assumes that the neutrino flux is
radial and the neutrino pressure tensor can be taken as
diagonal, thus ignoring effects due to neutrino viscosity. 
While the variable Eddington factor is determined as an
average value at all radii by solving the transport 
equations on an angularly averaged stellar background,
the multi-dimensionality of the problem is retained on
the level of the moment equations, which are radially
integrated within every angular zone of the spherical
coordinate grid. In addition, lateral gradients that
correspond to neutrino pressure and advection of
neutrinos with the moving stellar fluid are included in
the moment equations
(``ray-by-ray plus''). Note that neutrino pressure cannot
be ignored in the protoneutron star interior and advective
transport of neutrinos is faster than diffusion
below the neutrinosphere.

\begin{figure}[htp!]
\tabcolsep=0.5mm
\begin{tabular}{lr}
   \epsfxsize=0.49\hsize \epsfclipon \epsffile{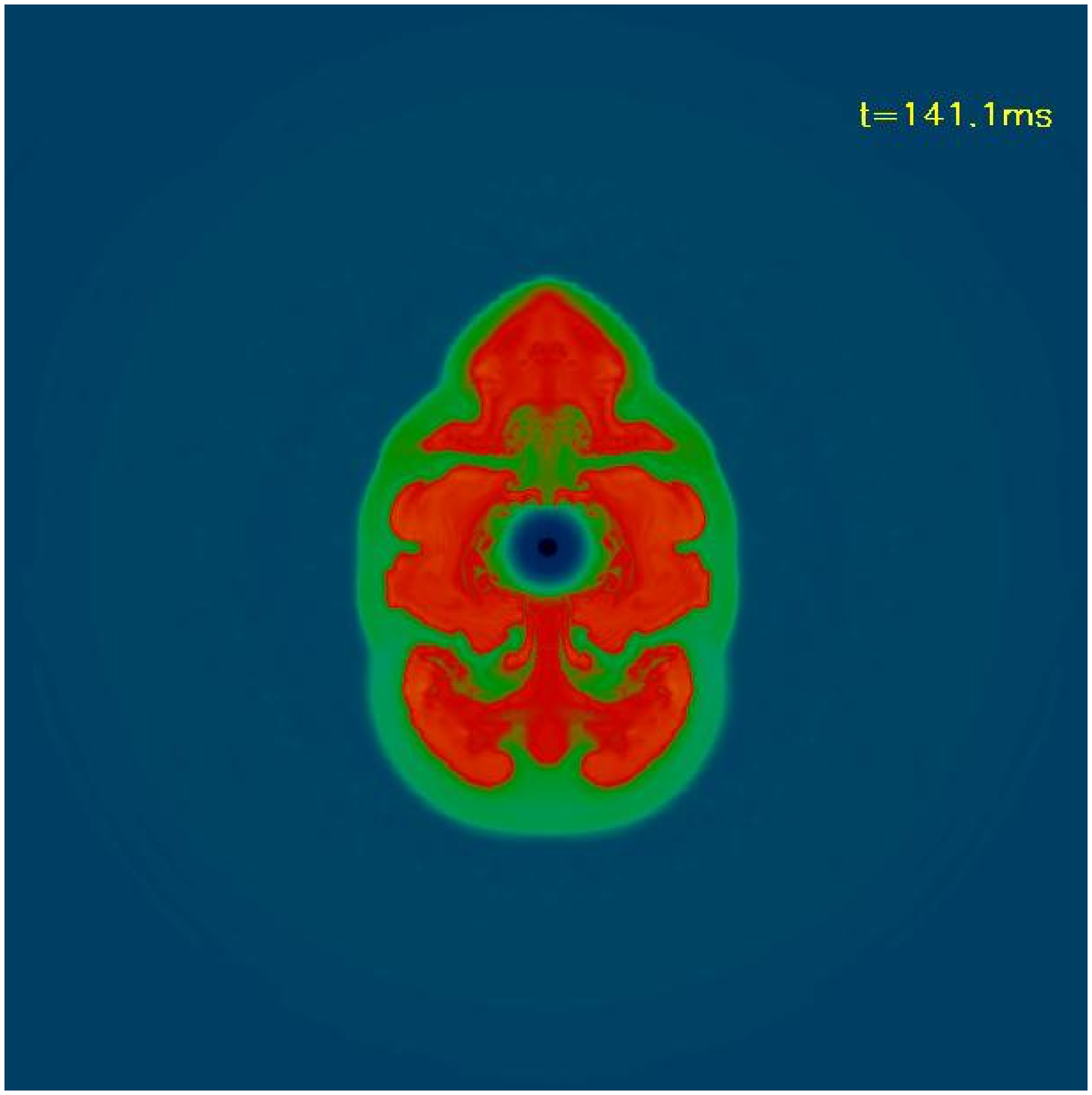} &
   \epsfxsize=0.49\hsize \epsfclipon \epsffile{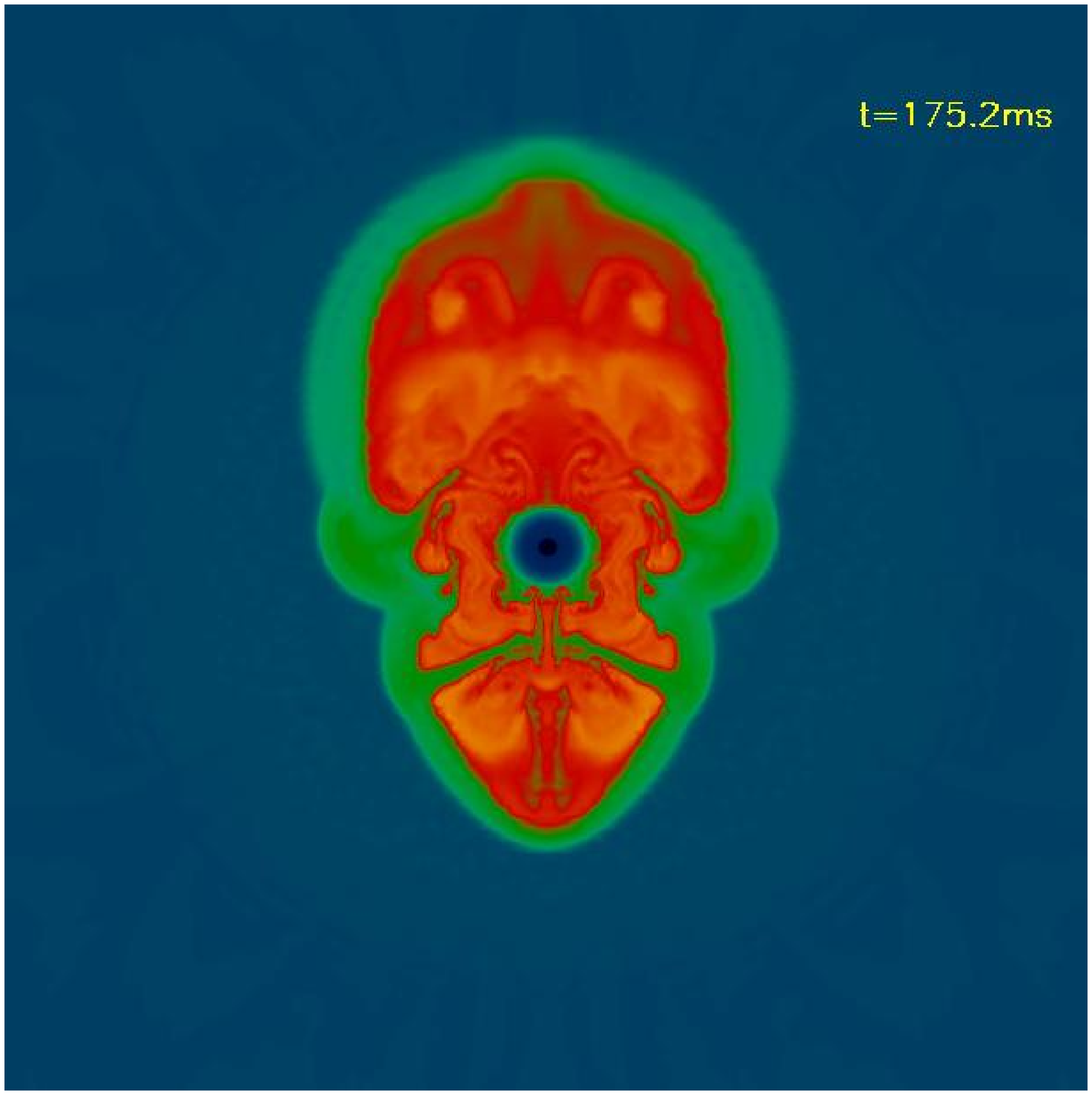} \\
   \epsfxsize=0.49\hsize \epsfclipon \epsffile{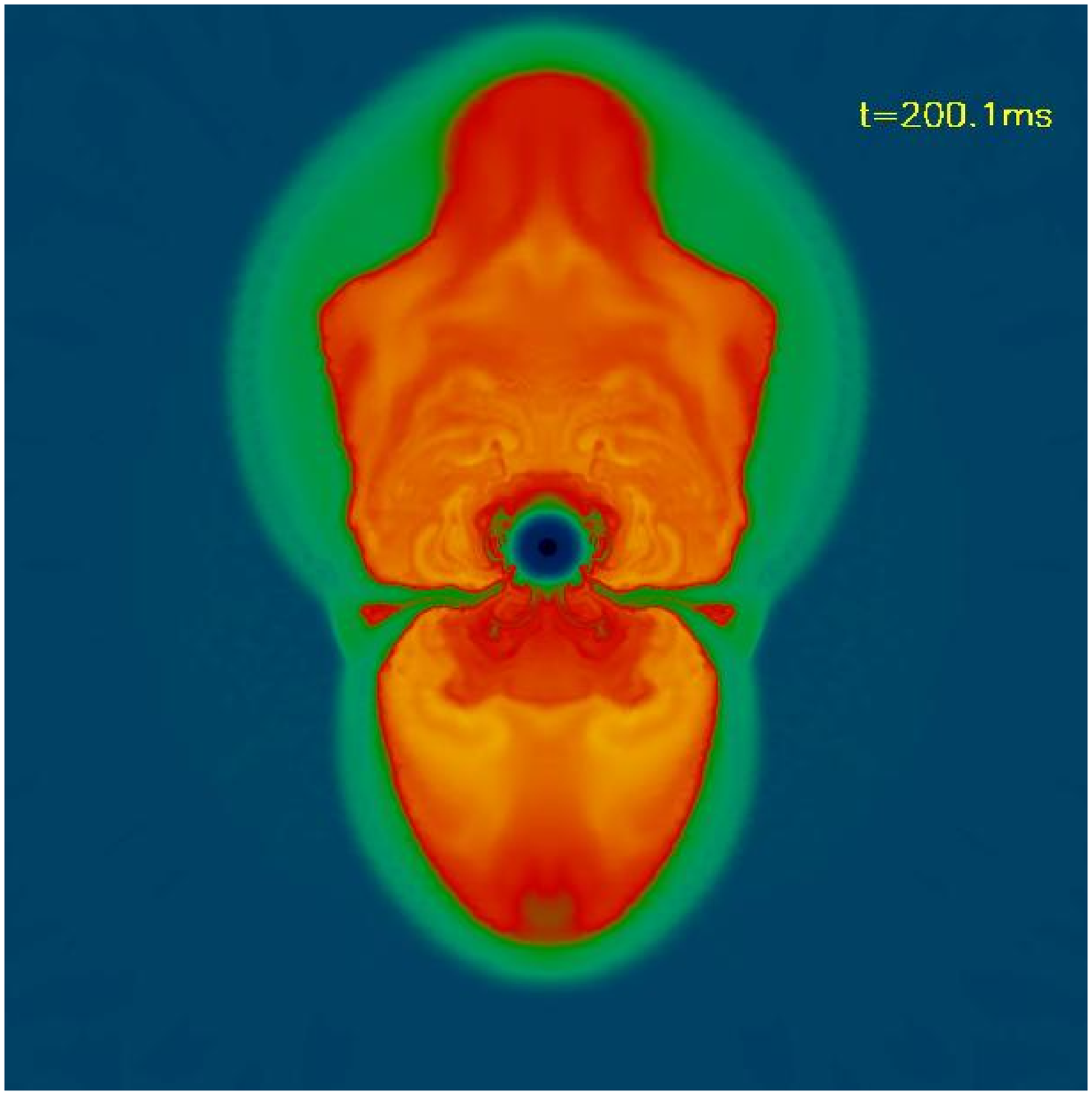} &
   \epsfxsize=0.49\hsize \epsfclipon \epsffile{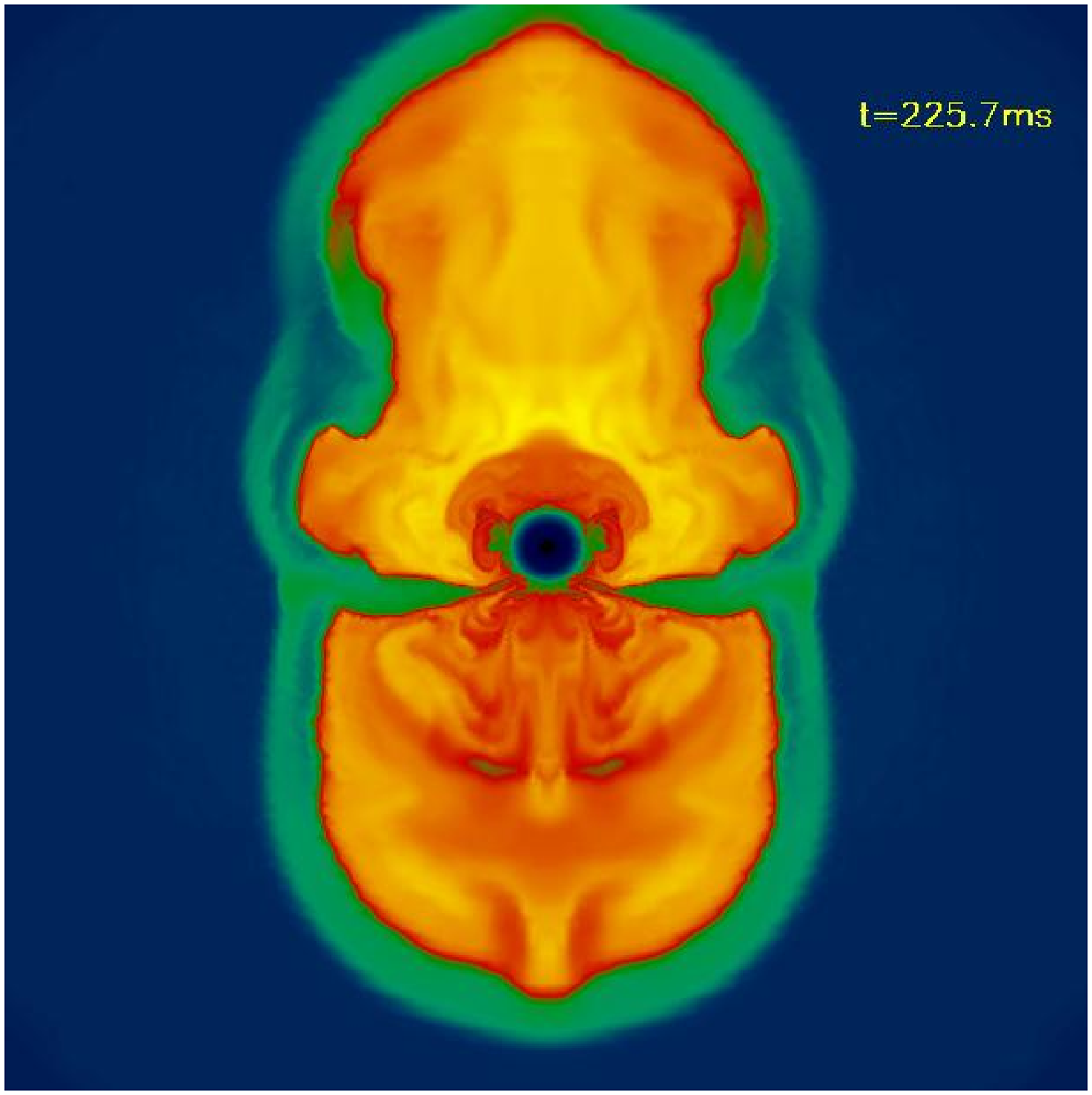} \\
\end{tabular}
\caption{
Sequence of snapshots showing the large-scale convective
overturn in the neutrino-heated postshock layer at four
post-bounce times ($t_{\mathrm{pb}} = 141.1\,$ms,
175.2$\,$ms, 200.1$\,$ms, and 225.7$\,$ms, from top left to
bottom right) during the evolution of a (non-rotating)
11.2$\,$M$_{\odot}$ progenitor model from Woosley et 
al.~\cite{thj_wooetal02}. 
The entropy is color coded with highest
values being represented by red and yellow, lowest
values by blue and black. The dense neutron star 
is visible as low-entropy circle at the center. The
convective layer interior of the neutrinosphere cannot be 
visualized with the employed color scale because the
entropy contrast there is small. Convection in this region
is driven by a negative gradient of the lepton number.
The computation was performed with 
spherical coordinates, assuming axial symmetry, and employing
the ``ray-by-ray plus'' variable Eddington factor technique
developed by Rampp \& Janka~\cite{thj_ramjan02} 
and Buras et al.~\cite{thj_buretal03}
for treating neutrino transport in multi-dimensional
supernova simulations. Equatorial
symmetry is broken on large scales soon after bounce, and
low-mode convection begins to dominate the flow between
the neutron star and the strongly deformed supernova shock.
The ``face'' on the top right does not need to look so sad,
because the model continues to develop a weak explosion.
The scale of the plots is 1200$\,$km in both directions.
}
\label{janka_fig1}
\end{figure}

Electron neutrinos and antineutrinos are produced by 
$e^-$ captures on nuclei and protons and $e^+$ captures
on neutrons, respectively. Nucleon-nucleon bremsstrahlung
and $e^+e^-$ annihilation are considered for the creation
of $\nu\bar\nu$ pairs of all flavors. Muon and tau
neutrino-antineutrino pairs are also made by $\nu_e\bar\nu_e$
annihilation. Neutrino scattering off $n$, $p$, $e^\pm$,
and nuclei is included, for muon and tau neutrinos also
off $\nu_e$ and $\bar\nu_e$. The charged-current reactions of
neutrinos with nucleons take into account nucleon thermal
motions, recoil and phase-space blocking, weak magnetism 
corrections, the reduction of the effective nucleon
mass and the quenching of the axial-vector coupling in 
nuclear matter, and nucleon correlations at high densities.

Recently, the treatment of electron captures on
heavy nuclei has been improved in collaboration with 
K.~Langanke and coworkers~\cite{thj_lanetal03}.
Details were reported at this meeting by 
G.~Mart\'{\i}nez-Pinedo. Previously these reactions
were described rather schematically~\cite{thj_bru85} and 
were switched off above a few $10^{10}\,$g$\,$cm$^{-3}$.
Therefore $e^-$ captures on $p$ determined the subsequent
evolution. In the new models, in contrast, nuclei dominate
the $\nu_e$ production by far. This leads to a significant
shrinking of the homologously collapsing inner core and
shock formation at a smaller mass coordinate~\cite{thj_lanetal03}.
Despite of this conceptually and quantitatively important
change the subsequent shock propagation and expansion 
remains astonishingly similar because of differential 
changes of the core structure and cancellations of 
effects~\cite{thj_hixetal03}.

\subsection{...for a new generation of multi-dimensional 
models...}

Running simulations for progenitors with different
main sequence masses (Woosley et al.'s 11.2, 15, and 
20$\,$M$_{\odot}$ models~\cite{thj_wooetal02}) 
in 1D and 2D, we could 
confirm the finding of previous multi-dimensional
models with simpler neutrino transport, namely that
two spatially separated regions exist in the supernova
core where convection sets in on a timescale of some
ten milliseconds after bounce~\cite{thj_buretal03}.

The one region is characterized by a negative entropy
gradient which is left behind by the weakening shock and
enhanced by the onset of neutrino heating between gain
radius and shock. Despite of a positive gradient of the
electron fraction, this region is Ledoux unstable and 
Rayleigh-Taylor mushrooms start to grow between 40$\,$ms
and 80$\,$ms post bounce (slower for higher-mass
progenitors). The violent convective overturn that
develops in this region supports the shock expansion
and allows for larger shock radii. Two effects seem to
be mainly responsible for this helpful influence on the
neutrino-heating mechanism. On the one hand bubbles of
neutrino-heated matter can rise, which pushes the shock 
farther out and reduces the energy loss by the reemission
of neutrinos. On the other hand, cold, lower-entropy 
matter is carried by narrow downflows from the shock to
near the gain radius, where it is heated by neutrinos at
very high rates. This enhances the efficiency of neutrino
energy transfer. Fully developed, the convective overturn
can become so violent that downflows penetrate with 
supersonic velocities through the electron neutrinosphere,
thereby increasing the luminosity of $\nu_e$ and $\bar\nu_e$.

The second region of convective activity lies beneath the 
neutrinosphere. Convection there is driven by a negative
lepton gradient and sets in between about 20$\,$ms post
bounce and about 60$\,$ms post bounce (again later for 
the more massive progenitors). Despite of the transport of
energy and lepton number and the corresponding change
of the outer layers of the protoneutron star,
the effect on the luminosities of $\nu_e$ and $\bar\nu_e$
is rather small. The neutrinosphere of heavy-lepton 
neutrinos, however, is located within the convective layer
and an enhancement of muon- and tauneutrino luminosities
(10--20\%) is visible at times somewhat later than
100$\,$ms. The influence 
on the shock propagation and the explosion mechanism is 
marginal and mostly indirect by modifying the neutron
star structure and $\nu_{\mu}$ and $\nu_\tau$ emission.

Although convective overturn behind the shock strongly
affects the post-bounce evolution, we were disappointed 
by not obtaining explosions in a recently published first 
set of simulations~\cite{thj_buretal03}. These results
seem to confirm the suspicion~\cite{thj_mezetal98}
that a more accurate treatment of neutrino
transport might not allow one to reproduce the convectively
supported neutrino-driven explosions seen previously.

\subsection{...with ultimate success?}

But there is light at the end of the long tunnel and the
situation is more favorable than it looks at first glance.
There are reasons to believe that our models are very
close to explosions, in fact graze the threshold of
conditions which are required to drive mass ejection 
by the outward acceleration of the supernova shock.

One of our models (a 15$\,$M$_{\odot}$ star) included 
rotation at a rate that is consistent with pre-collapse core
rotation of magnetized stars~\cite{thj_hegetal03}. 
The assumed initial
angular velocity was chosen to be slightly faster 
($\Omega = 0.5\,$rad$\,$s$^{-1}$) than predicted
by Heger et al.~\cite{thj_hegetal03}. It would lead to
a neutron star spinning with a period of 1--2$\,$ms if 
the angular momentum in the protoneutron star is conserved
after the end of our simulations.
We intentionally did not consider more extreme rotation
rates which are expected for collapsars and needed for 
gamma-ray burst models, but which are probably not generic 
for supernovae. 

Rotation makes a big difference! Centrifugal forces 
reduce the infall velocity near the equatorial plane 
and help to support the shock at a larger radius.
Enhanced by rotationally induced vortex motion
extremely violent convective overturn develops behind
the shock. Powerful non-radial oscillations
are initiated and drive the shock temporarily to distances
near 300$\,$km along the rotation axis where the more 
rapidly decreasing density favors strong shock expansion.

Huge global deformation was also observed in case of the 
(non-rotating) 11.2$\,$M$_{\odot}$ star when we increased 
the angular
grid from a $\sim$90$^{\mathrm o}$ wedge ($\pm 46.8^{\mathrm o}$
around the equatorial plane of the coordinate grid
with periodic boundary conditions) to full 180$^{\mathrm o}$.
The 11.2$\,$M$_{\odot}$ model is characterized by a small iron
core ($\sim$1.25$\,$M$_{\odot}$) and an abrupt entropy jump
at the edge of the Si shell (at $\sim$1.3$\,$M$_{\odot}$).
A strong dipolar expansion occurs and the shock is slowly 
pushed outward by the pulsational expansion of two huge
bubbles which are alternately fed by neutrino heated matter
that comes from a single (due to the assumed symmetry, 
toroidal), waving downflow near the equatorial
plane of the coordinate grid (Fig.~\ref{janka_fig1}). 
The shock has reached a maximum radius of more than 600$\,$km
with no sign of return until we had to stop the simulation
226$\,$ms after bounce.

\begin{figure}[htp!]
\tabcolsep=1.5mm
\begin{tabular}{lr}
   \epsfysize=0.45\hsize \epsfclipon \epsffile{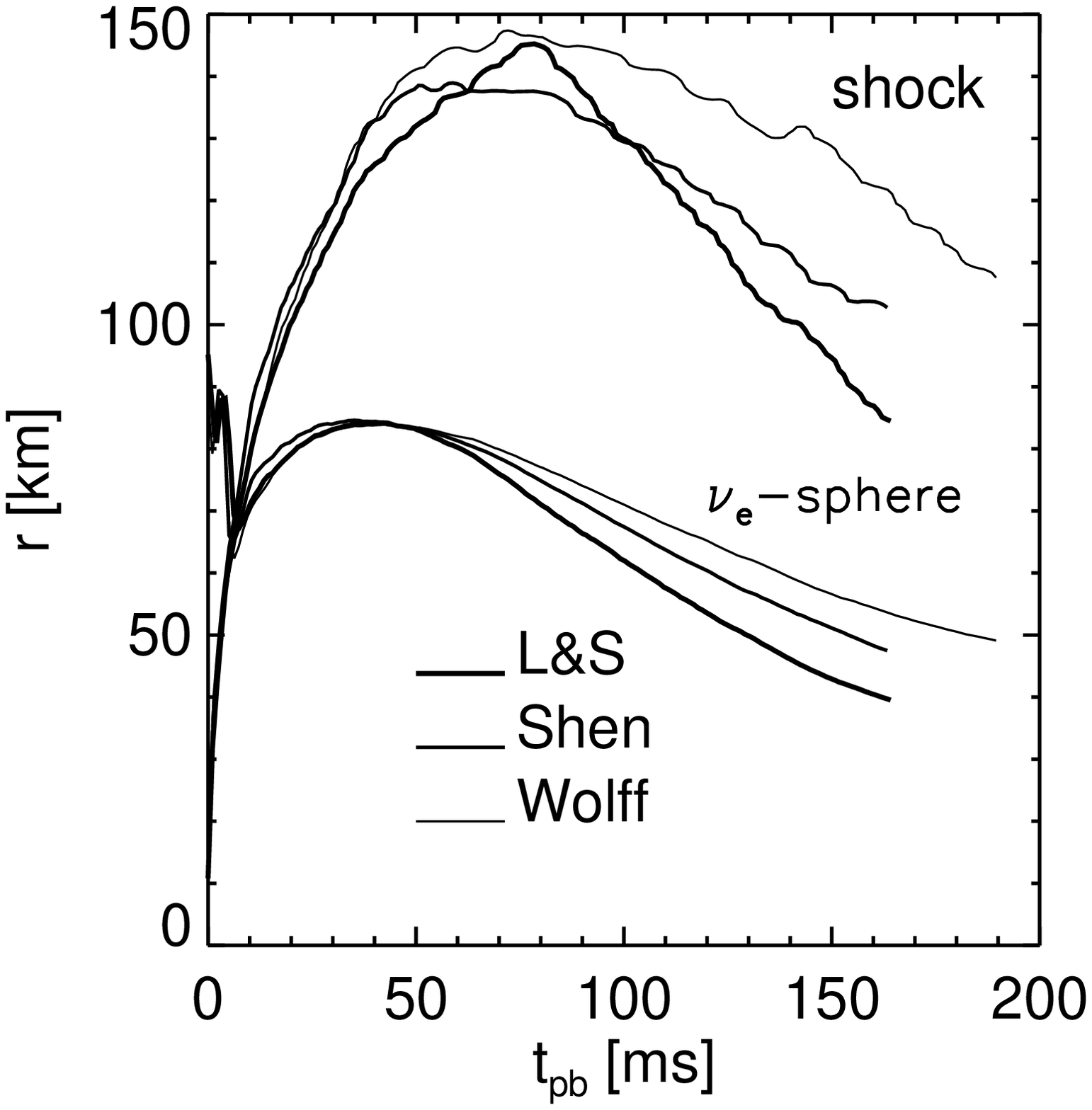} &
   \epsfysize=0.45\hsize \epsfclipon \epsffile{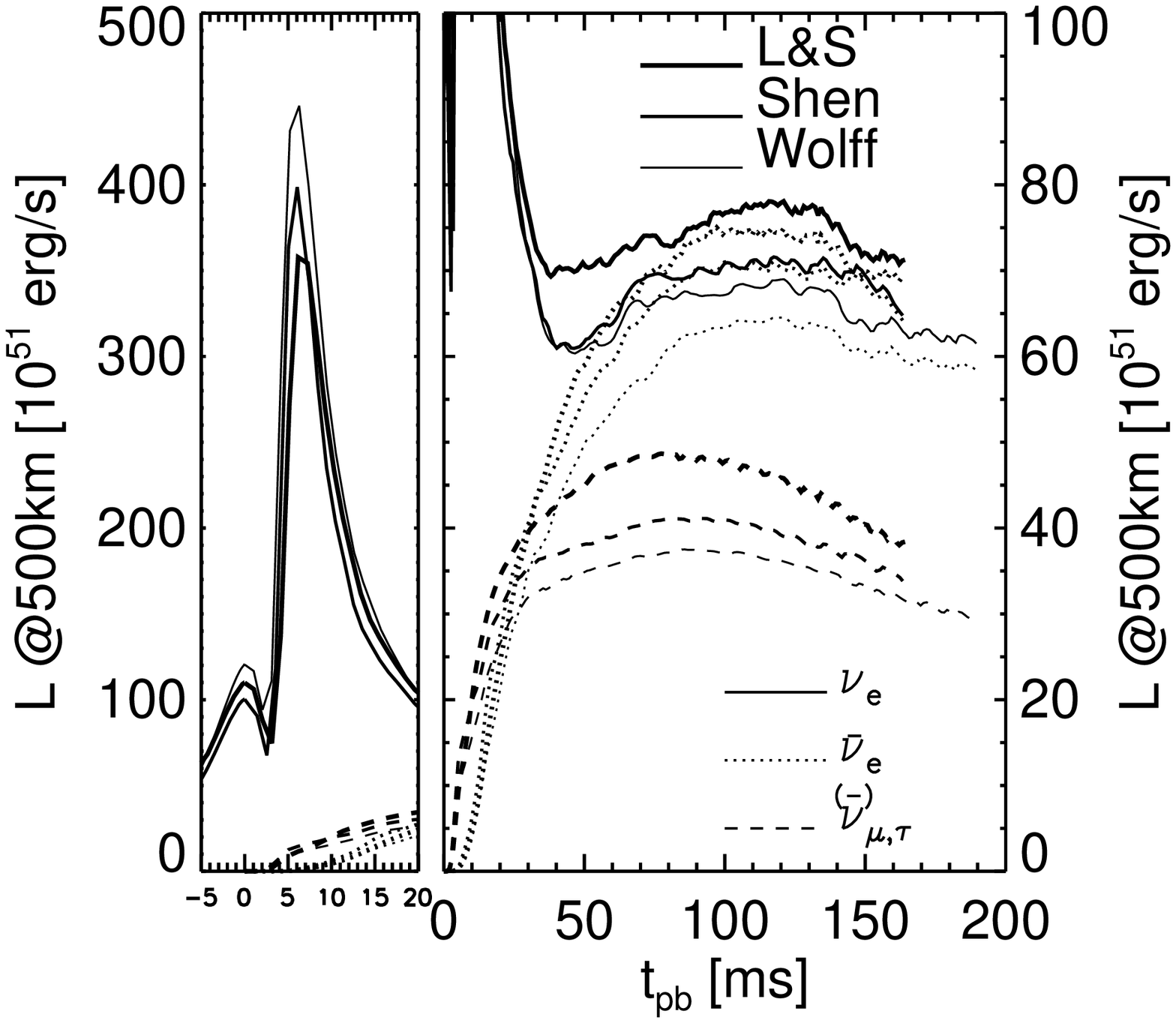} \\
   \parbox[t]{0.43\hsize}{\caption[]{
Shock radii and electron neutrinospheres for   
simulations of a 15$\,$M$_{\odot}$ star in spherical symmetry with
three different nuclear EoSs~\cite{thj_marek03}, namely those
of Lattimer \& Swesty (\cite{thj_latswe91}; bold lines),
which is the widely
used standard for supernova simulations these days,
Shen et al. (\cite{thj_shenetal}; medium),
and Wolff \& Hillebrandt (\cite{thj_hilwol85}; thin).
Times are synchronized at the moment of core bounce.
}
\label{janka_fig2}} &
    \parbox[t]{0.51\hsize}{\caption[]{
Luminosities for $\nu_e$, $\bar\nu_e$, and heavy-lepton neutrinos
($\nu_{\mu}$, $\nu_{\tau}$, $\bar\nu_{\mu}$ or $\bar\nu_{\tau}$
individually), measured by an observer comoving with the
infalling stellar plasma at a radius of 500$\,$km, for the three
spherically symmetric simulations shown in Fig.~\ref{janka_fig2}
\cite{thj_marek03}. The left panel displays a time interval around
the prompt $\nu_e$ burst, the right panel a longer period of the
post-bounce evolution. Note the different scales on the vertical
axes of both frames.
}
\label{janka_fig3}} \\
\end{tabular}
\end{figure}

We consider it as very likely that a weak explosion develops
in this model. It is exciting to imagine how the evolution
might have proceeded with the additional help from rotation.
Patience, however, is necessary when results for longer
post-bounce periods or other progenitors are desired.
The computations require far more than $10^{17}$ floating
point operations and take several months on machines
available to us.

We actually have hints of how an explosion can emerge
in a 15$\,$M$_{\odot}$ star which was computed with
omitted velocity-dependent terms in the neutrino
momentum equation. The resulting 20--30\% change
of the neutrino density between neutrinosphere and shock 
was sufficient to initiate an explosion, thus
demonstrating that not much was missing 
for the convectively supported neutrino-heating
mechanism to work. The explosion had an energy of 
about $6\times 10^{50}\,$erg and left behind a neutron star
with an initial baryonic mass of $\sim$1.4$\,$M$_{\odot}$. The
neutrino-heated ejecta did not show the dramatic
overproduction of $N=50$ closed neutron shell nuclei which
signaled a problem with the neutrino transport
approximations used in previous models. 

These results suggest that we are 
on the right track. Once the critical threshold
for explosions can be overcome, the subsequent evolution
seems to proceed very favorably with respect to observable
facts.

\section{Longing for more}
\label{janka_sec:4}
What can provide or support the ultimate kick beyond the explosion
threshold? Is it three-dimensional effects? Very fast rotation?
Truely multi-dimensional transport that accounts for lateral 
neutrino flow and neutrino shear? Full general relativity
instead of approximations? 
Or yet to be improved microphysics, e.g. reactions of 
neutrinos with nuclei? Or the uncertain high-density equation 
of state which has not been extensively varied up to now
but can cause sizable differences 
(Figs.~\ref{janka_fig2},\ref{janka_fig3})? Or so far
ignored or unresolved modes of instability that could boost 
the neutrino luminosity or drive accretion shock instability?
Or magnetohydrodynamic effects?
Or is it the combination of all?

Much work still needs to be done for completing the 
supernova codes and testing these possibilities. A number
of groups around the world have set out to meet this challenge!

\bigskip
{\small
\noindent
{\bf Acknowledgements.}
Support by the Sonderforschungsbereich SFB-375 
``Astro-Teil\-chen\-phy\-sik'' of the Deutsche Forschungsgemeinschaft
is acknowledged. The simulations were done on the IBM ``Regatta''   
supercomputer of the Rechenzentrum Garching.
}

%
{\small
\input{janka_ref}

}


\end{document}

%% file: janka_ref.tex
%
%

%
%